\documentclass[intlimits,twoside,a4paper]{article}

\usepackage{amsmath,amssymb}
\usepackage{graphicx}
\usepackage{epsfig}

\usepackage[T2A]{fontenc}
\usepackage[cp1251]{inputenc}

\usepackage{cmpj2}
\issue{2014}{17}{4}{43003}
\doinumber{10.5488/CMP.17.43003}

\title{Random-field Ising model: Insight from zero-temperature simulations}

\author[P.E. Theodorakis, N.G. Fytas]{P.E. Theodorakis\refaddr{ad1}, N.G. Fytas\refaddr{ad2}}

\addresses{\addr{ad1}Department of Chemical Engineering, Imperial
College London, SW7 2AZ, London, United Kingdom
\addr{ad2}Applied Mathematics Research Centre, Coventry
University, Coventry, CV1 5FB, United Kingdom}

\date{Received October 2, 2014}
\authorcopyright{P.E. Theodorakis, N.G. Fytas, 2014}

\begin{document}

\maketitle

\begin{abstract}
We enlighten some critical aspects of the three-dimensional
($d=3$) random-field Ising model from simulations performed
at zero temperature. We consider two different, in terms of the
field distribution, versions of model, namely a Gaussian random-field Ising model and
an equal-weight trimodal random-field Ising model. By implementing a computational
approach that maps the ground-state of the system to the
maximum-flow optimization problem of a network, we employ the most
up-to-date version of the push-relabel algorithm and simulate
large ensembles of disorder realizations of both models for a
broad range of random-field values and systems sizes
$\mathcal{V}=L\times L\times L$, where $L$ denotes linear lattice
size and $L_{\rm max}=156$. Using as finite-size measures the
sample-to-sample fluctuations of various quantities of physical
and technical origin, and the primitive operations of the
push-relabel algorithm, we propose, for both types of
distributions, estimates of the critical field $h_{\rm c}$ and the
critical exponent $\nu$ of the correlation length, the latter
clearly suggesting that both models share the same universality
class. Additional simulations of the Gaussian random-field Ising model at the
best-known value of the critical field provide the magnetic
exponent ratio $\beta/\nu$ with high accuracy and clear out the
controversial issue of the critical exponent $\alpha$ of the
specific heat. Finally, we discuss the infinite-limit size
extrapolation of energy- and order-parameter-based noise to signal
ratios related to the self-averaging properties of the model, as
well as the critical slowing down aspects of the algorithm.
\end{abstract}

\section{Introduction}
\label{sec:1}

The random-field Ising model (RFIM) is one of the archetypal disordered
systems~\cite{imr75,aharony76,parisi79}, extensively studied due
to its theoretical interest, as well as its close connection to
experiments in hard~\cite{by1991,rieger1995} and soft condensed
matter systems~\cite{vink}. Its beauty is that the mixture of
random fields and the standard Ising model creates rich physics
and leaves many still unanswered problems. The Hamiltonian
describing the model is
\begin{equation}
\label{eq:1}
\mathcal{H}=-J\sum_{\langle i,j \rangle}\sigma_{i}\sigma_{j}-\sum_{i}h_{i}\sigma_{i}\,,
\end{equation}
where $\sigma_{i}=\pm 1$ are Ising spins, $J>0$ is the
nearest-neighbor's ferromagnetic interaction, and $h_{i}$ are
independent quenched random fields.

\begin{figure}[!t]
\vspace{-5mm}
\centerline{
\includegraphics*[width=0.5\textwidth]{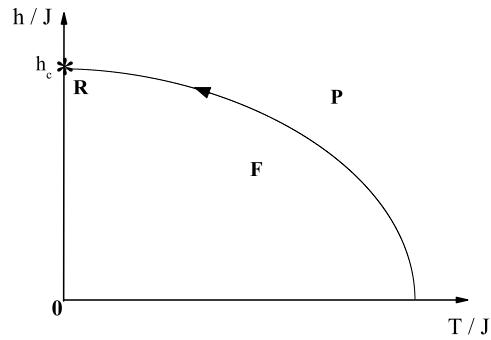}
}
\caption{\label{fig:1} Schematic phase diagram and
renormalization-group flow of the RFIM. The solid line separates
the ferromagnetic (\textbf{F}) and paramagnetic (\textbf{P})
phases. The black arrow shows the flow to the random fixed point
(\textbf{R}) at $T=0$ and $h=h_{\rm c}$, as marked by an
asterisk.}
\end{figure}

The existence of an ordered ferromagnetic phase for the RFIM, at
low temperature and weak disorder, followed from the seminal
discussion of Imry and Ma~\cite{imr75}, when the space dimension
is greater than two
($d>2$)~\cite{villain84,bray85,fisher86,berker86,bricmont87}. This
has provided us with a general qualitative agreement on the sketch
of the phase boundary, separating the ordered ferromagnetic phase
from the high-temperature paramagnetic one. The phase-diagram line
separates the two phases of the model and intersects the
randomness axis at the critical value of the disorder strength
$h_{\rm c}$, as shown in figure~\ref{fig:1}. Such qualitative
sketching has been commonly used for the
RFIM~\cite{newman93,machta00,barkema} and closed form quantitative
expressions are also known from the early mean-field
calculations~\cite{aharony78,aharony78.2,aharony78.3}. However, it is generally true that
the quantitative aspects of phase diagrams produced by mean-field
treatments provide rather poor approximations.

The criteria for determining the order of the low-temperature
phase transition and its dependence on the form of the field
distribution have been discussed throughout the
years~\cite{aharony78,aharony78.2,aharony78.3,galam,saxena84}. Although the view that the
phase transition of the RFIM is nowadays considered to be of
second
order~\cite{hartmann-99,middleton1,vink10,yllanes,fytasme,fytas2},
the extremely small value of the exponent $\beta$ continues to
cast some doubts. Moreover, a rather strong debate with regard to
the role of disorder, i.e., the dependence, or not, of the
critical exponents on the particular choice of the distribution
for the random fields and the value of the disorder strength,
analogously to the mean-field theory predictions \cite{aharony78,aharony78.2,aharony78.3},
was only recently put on a different basis \cite{fytasPRL}.
Currently, even the well-known correspondence among the RFIM and
its experimental analogue, the diluted antifferomagnet in a field
(DAFF), has been severely questioned by extensive simulations
performed on both models at positive- and
zero-temperature~\cite{ahrensDAFF}. In any case, the whole issue
of the model's critical behavior is still under intense
investigation~\cite{hartmann-99,middleton1,vink10,yllanes,
fytasme,fytas2,rieger93,rieger93.2,falicov95,swift97,sourlas,sourlas',nowak98,duxbury01,ceva,crokidakis,crokidakis',giannis,giannis.2,giannis.3,akinci,tarjus}.

Already from the work of Houghton {et al.}~\cite{houghton85},
the importance of the form of the distribution function in the
determination of the critical properties of the RFIM has been
emphasized. In fact, different results have been proposed for
different field distributions, like the existence of a tricritical
point at the strong disorder regime of the system, present only in
the bimodal case \cite{aharony78,aharony78.2,aharony78.3,houghton85}. Following the
results of Houghton {et al.} \cite{houghton85},
Mattis \cite{mattis} reexamined the RFIM introducing a new type of
a trimodal distribution
\begin{equation}
\label{eq:2} \mathcal{P}^{\rm
(trimodal)}(h_{i})=p\delta(h_{i})+\left(\frac{1-p}{2}\right)[\delta(h_{i}-h)+\delta(h_{i}+h)],
\end{equation}
where $h$ defines the disorder (field) strength and $p\in (0,1)$.
Clearly, for $p=1$ one switches to the pure Ising model, whereas
for $p=0$ the well-known bimodal distribution is recovered. In
general terms, the trimodal distribution~(\ref{eq:2}) permits a
physical interpretation as a diluted bimodal distribution, in
which a fraction $p$ of the spins are not exposed to the external
field. Thus, it mimics the salient feature of the Gaussian
distribution
\begin{equation}
\label{eq:3} \mathcal{P}^{\rm
(Gaussian)}(h_{i})=\frac{1}{\sqrt{2\pi
h^{2}}}\exp{\left(-\frac{h_{i}^{2}}{2h^{2}}\right)},
\end{equation}
for which a significant fraction of the spins are in weak external
fields. Mattis suggested that for a particular case, $p=1/3$,
equation (\ref{eq:2}) may be considered to a good approximation as the
Gaussian distribution~\cite{mattis}. This in turn indicated that
the two models should be in the same universality class. Further
studies along these lines, using mean-field and
renormalization-group approaches, provided contradicting evidence
for the critical aspects of the $p=1/3$ model and also proposed
several approximations of its phase diagram for a range of values
of $p$~\cite{kaufman,sebastianes,arruda}. However, none of these
predictions has been confirmed by numerical simulations up till now,
thus remaining ambiguous, due to the approximate nature of the
mean-field-type of the methods used.

The scope of the present work is to shed some light towards this
direction by examining several critical features of the phase
diagram of the RFIM at $d=3$, using both distributions described
above in equations~(\ref{eq:2}) and (\ref{eq:3}). In particular, in the
first part of our study we provide numerical evidence that clarify
the matching between the trimodal ($p=1/3$) and Gaussian models
and we give estimates for the critical field $h_{\rm c}$ and
critical exponent $\nu$ that compare very well to the most
accurate ones in the corresponding literature of the RFIM. In the
second part of our study we concentrate on the most studied case
of the Gaussian RFIM, for which we present a scaling analysis of
critical data for the order parameter and the specific heat, i.e.,
data obtained at the best known estimate of the critical field
$h_{\rm c}$. Our analysis points to a very small, but non-zero,
value for the magnetic exponent ratio $\beta /\nu$, and a critical
exponent $\alpha \rightarrow 0^{-}$, in good agreement with
experimental predictions~\cite{jaccarino,jaccarino.2}. We also discuss the
infinite-limit size extrapolation of energy- and
order-parameter-based noise to signal ratios related to the
self-averaging properties of the model, as well as some technical
aspects of the implemented numerical method.

Our attempt benefits from: (i) the existence of robust
computational methods of graph theory at zero temperature ($T=0$)
that allow us to simulate very large system sizes and disorder
ensembles, necessary for an accurate investigation of the delicate
properties discussed above, (ii) classical finite-size scaling
(FSS) techniques, and (iii) a new scaling approach that involves
the analysis of the sample-to-sample fluctuations of various
well-defined quantities. In particular, sample-to-sample
fluctuations and the relative issue of self-averaging have
attracted much interest in the study of disordered
systems~\cite{brout59}. Although it has been known for many years
now that for (spin and regular) glasses there is no self-averaging
in the ordered phase~\cite{binder86}, for random ferromagnets such
a behavior was first observed for the RFIM by Dayan {et al.}~\cite{dayan93} and some years later for the random versions
of the Ising and Ashkin-Teller models by Wiseman and
Domany~\cite{WD95}. These latter authors suggested a FSS ansatz
describing the absence of self-averaging and the universal
fluctuations of random systems near critical points that was
refined on a more rigorous basis by Aharony and
Harris~\cite{AH96}. Ever since, the subject of breakdown of
self-averaging is an important aspect of several theoretical and
numerical investigations of disordered spin
systems~\cite{EB96,PSZ97,WD98,BF98,TO01,PS02,BC04,MG05,fytas06,wu06,wu06',GL07,fytas10}.
In fact, Efrat and Schwartz~\cite{efrat06} showed that the
property of lack of self-averaging may be turned into a useful
tool that can provide an independent measure to distinguish the
ordered and disordered phases of the system. In view of this
observation, we discuss here another useful application of the
fluctuation properties of several quantities of the system to
obtain information on the ground-state criticality of the RFIM.

The rest of the paper is organized as follows: In the next section
we describe the general framework behind the mapping of the RFIM
to the corresponding network, outline the numerical approach, and
provide all the necessary details of our implementation. The
relevant FSS analysis that shows the equivalence of both
distributions under study in terms of the critical exponent $\nu$
of the correlation length, using an approach based on the
sample-to-sample fluctuations of the model, is presented in
section~\ref{sec:3}. Then, in section~\ref{sec:4} we focus our attention
on the most studied case of the Gaussian model and we provide
estimates for the magnetic exponent ratio $\beta/\nu$ and the
critical exponent $\alpha$ of the specific heat, via the scaling
of the order parameter and bond energy, respectively, at the best
known estimate of the critical field value. We also investigate
the self-averaging properties of the model at criticality, using
properly defined noise to signal ratios and we provide an estimate
for the exponent $z$ that describes the critical slowing of the
algorithm used. Finally, we synopsize our findings in
section~\ref{sec:5}.

\section{Simulation protocol} \label{sec:2}

As already discussed extensively in the literature (see
reference~\cite{hartmannbook1} and references therein), the RFIM
captures essential features of models in statistical physics that
are controlled by disorder and have frustration. Such systems show
complex energy landscapes due to the presence of large barriers
that separate several meta-stable states. When such models are
studied using simulations mimicking the local dynamics of physical
processes, it takes an extremely long time to encounter the exact
ground state. However, there are cases where efficient methods for
finding the ground state can be utilized and, fortunately, the
RFIM is one such case. These methods escape from the typical
direct physical representation of the system, in a way that extra
degrees of freedom are introduced and an expanded problem is
finally solved. By expanding the configuration space and choosing
proper dynamics, the algorithm practically avoids the need of
overcoming large barriers that exist in the original physical
configuration space. An attractor state in the expended space is
found in time polynomial in the size of the system and when the
algorithm terminates, the relevant auxiliary fields can be
projected onto a physical configuration, which is the guaranteed
ground state.

The random field is a relevant perturbation at the pure
fixed-point, and the random-field fixed-point is at
$T=0$~\cite{villain84,bray85,fisher86,berker86}. Hence, the
critical behavior is the same everywhere along the phase boundary
of figure~\ref{fig:1}, and we can predict it simply by staying at
$T=0$ and crossing the phase boundary at $h=h_{\rm c}$. This is a
convenient approach, because we can determine the ground states of
the system exactly using efficient optimization
algorithms~\cite{hartmann-99,middleton1,fytas2,wu06,wu06',ogielski85,hartmann01,machta03,seppala,alava,fytas}
through an existing mapping of the ground state to the
maximum-flow optimization problem~\cite{papadimitriou}. A clear
advantage of this approach is the ability to simulate large system
sizes and disorder ensembles in rather moderate computational
times. We should underline here that, even the most efficient
$T>0$ Monte Carlo schemes exhibit extremely slow dynamics in the
low-temperature phase of these systems and are upper bounded by
linear sizes of the order of $L_{\rm max}\leqslant
32$~\cite{hartmannbook1}. Further advantages of the $T=0$ approach
are the absence of statistical errors and equilibration problems,
which, on the contrary, are the two major drawbacks encountered in
the $T>0$ simulation of systems with rough free-energy
landscapes~\cite{rieger1995}.

A short direct sketching of how this mapping may in principle
occur through some simple considerations is as follows: Let
$G=(V,E)$ be a directed, weighted graph consisting of a set $V$ of
nodes and a set $E$ of edges, each of the latter connecting two
nodes. In a directed graph, for each edge a direction is
specified. The property of being weighted means that to each edge
from node $i$ to node $j$ a capacity $c_{ij}$ is assigned. Let the
number of nodes be $n+2$. We enumerate the nodes
$V=\{0,1,2,\ldots,n,n+1\}$ and define the first node 0 as source
$s$ and the last node $n+1$ as the sink $t$. The remaining nodes
will be associated to the lattice sites of the RFIM. We call a
directed, weighted graph $G$ with source $s$, sink $t$, and
capacities $c$, as network $\mathcal{N}=(G,c,s,t)$. Now, in a
network $\mathcal{N}=(G,c,s,t)$, an ($s,t$)-cut ($S,\overline{S}$)
is defined as a partition of the set of nodes $V$ into two
disjoint sets $S$ and $\overline{S}$ ($S \cap
\overline{S}=\varnothing$ and $ S \cup \overline{S}=V$) with $s\in
S$ and $t\in \overline{S}$. In other words, one can imagine a cut
as a partition that divides the network into two parts, one part
belonging to the source and the other to the sink. Generally,
there are many different possible cuts in a network. We can assign
to each of them a capacity $C(S,\overline{S})$, defined as the sum
of the capacities of the edges that the cut crosses
\begin{equation}
\label{eq:4} C(S,\overline{S})=\sum_{i\in S, j\in \overline{S}}
c_{ij}\,.
\end{equation}
Note that edges are directed, that is why only edges that start at the
source side of the cut and end at the sink side contribute to the
capacity of the cut.

Now, the central idea that allows us to map the RFIM into a
network defined above, consists of describing a cut by a vector
$X$ with the property: $X_{i}=1$ if $i\in S$ and $X_{i}=0$
otherwise, i.e., if $i\in \overline{S}$. Then, by definition,
$X_{0}=1$ and $X_{n+1}=0$. Using this representation, the formula
for the cut capacity may be written in the following form
\begin{equation}
\label{eq:5}
C(S,\overline{S})=\sum_{i=0}^{n+1}\sum_{j=0}^{n+1}c_{ij}X_{i}(1-X_{j}).
\end{equation}
An expansion of equation~(\ref{eq:5}) leads to
\begin{equation}
\label{eq:6}
C(S,\overline{S})=-\sum_{i,j}c_{ij}X_{i}X_{j}+\sum_{i}\left(\sum_{j}c_{ij}\right)
X_{i}\,,
\end{equation}
and already a structural similarity to the fundamental Hamiltonian
definition of the RFIM [equation~(\ref{eq:1})] is clearly seen. Further
information on this structural similarities, including a detailed
algebra, may be found for the interested reader in the relevant
literature (see for instance reference~\cite{hartmannbook1} and
references therein).

The application of maximum-flow algorithms to the RFIM is nowadays
well established~\cite{alava}. The most efficient network flow
algorithm used to solve the RFIM is the push-relabel (PR)
algorithm of Tarjan and Goldberg~\cite{tarjan}. For the interested
reader, general proofs and theorems on the PR algorithm can be
found in standard textbooks~\cite{papadimitriou}. The version of
the algorithm implemented in our study involves a modification
proposed by Middleton {et al.}~\cite{middleton1,middleton2,middleton3} that removes the
source and sink nodes, reducing memory usage and also clarifying
the physical connection~\cite{middleton2,middleton3}.

The algorithm starts by assigning an excess $x_i$ to each lattice
site $i$, with $x_i = h_i$. Residual capacity variables $r_{ij}$
between neighboring sites are initially set to $J$. A height
variable $u_i$ is then assigned to each node via a global update
step. In this global update, the value of $u_i$ at each site in
the set ${\cal T} =\left\{j|x_j<0\right\}$ of negative excess
sites is set to zero. Sites with $x_i \geqslant 0$ have $u_i$ set to the
length of the shortest path, via edges with positive capacity,
from $i$ to ${\cal T}$. The ground state is found by successively
rearranging the excesses $x_i$, via \emph{push} operations, and
updating the heights, via \emph{relabel} operations. The order in
which sites are considered is given by a queue. In this paper, we
consider a first-in-first-out (FIFO) queue. The FIFO structure
executes a PR step for the site $i$ at the front of a list. If any
neighboring site is made active by the PR step, it is added to the
end of the list. If $i$ is still active after the PR step, it is
also added to the end of the list. This structure maintains the
cycles through the set of active sites.

When no more pushes or relabels are possible, a final global
update determines the ground state, so that sites which are path
connected by bonds with $r_{ij}>0$ to ${\cal T}$ have
$\sigma_i=-1$, while those which are disconnected from ${\cal T}$
have $\sigma_i = 1$. A push operation moves excess from a site $i$
to a lower height neighbor $j$, if possible, that is, whenever
$x_i>0$, $r_{ij} > 0$, and $u_j = u_i-1$. In a push, the working
variables are modified according to $x_i \rightarrow x_i -
\delta$, $x_j \rightarrow x_j + \delta$, $r_{ij} \rightarrow
r_{ij} - \delta$, and $r_{ji} \rightarrow r_{ji} + \delta$, with
$\delta = \min(x_i, r_{ij})$. Push operations tend to move the
positive excess towards sites in ${\cal T}$. When $x_i > 0$ and no
push is possible, the site is relabelled, with $u_i$ increased to
$1 + \min_{\{j| r_{ij} > 0\}} u_j$. In addition, if a set of
highest sites ${\cal U}$ become isolated, with $u_i > u_j+1$, for
all $i\in{\cal U}$ and all $j\notin{\cal U}$, the height $u_i$ for
all $i\in{\cal U}$ is increased to its maximum value,
$\mathcal{V}$, as these sites will always be isolated from the
negative excess nodes.

Periodic global updates are often crucial to the practical speed
of the algorithm \cite{middleton2,middleton3}. Following the
suggestions of references \cite{middleton1,middleton2,middleton3}, we
have also applied global updates here every $\mathcal{V}$
relabels, a practice found to be computationally
optimum \cite{fytas2,fytas,middleton2,middleton3}.

Using this scheme we performed large-scale simulations of the RFIM
with both type of distributions discussed above in
section~\ref{sec:1}. Let us note here that prior to the commencement
of these large-scale simulations, a set of preliminary runs with
smaller system sizes revealed the critical $h$-regime that we
should work on. In particular, for the trimodal ($p=1/3$) RFIM
simulations have been performed for lattice sizes
$L\in\{24,32,48,64,96,128\}$ and disorder strengths $h\in [2.7 -
3.3]$. For the Gaussian model lattice sizes in the range
$L\in\{L_{\rm min} - L_{\rm max}\}$, where $L_{\rm min}=24$ and
$L_{\rm max}=156$, were used and disorder strengths $h\in [2.0 -
3.0]$. In both cases a disorder-strength step of $\delta h=0.02$
was used. Regarding the disorder averaging procedure, which is of
paramount importance in the study of the RFIM, for each pair
$(L,\;h)$ an extensive averaging over $\mathcal{N}_{\rm
s}=50\times 10^3$ independent random-field realizations has been
undertaken, much larger than in previous relevant studies of the
model~\cite{middleton1,wu06,wu06',hartmann01,machta03}. Additionally,
for the Gaussian RFIM we performed some further and even more
extensive simulations, at the best known estimate of the critical
field $h_{\rm c}$, using in this case an ensemble of
$\mathcal{N}_{\rm s}=200\times 10^3$ random realizations.

\section{Universality aspects}
\label{sec:3}

As the outcome of the PR algorithm is the spin configuration of
the ground state, we can calculate for a given sample of a lattice
with linear size $L$ the magnetization via $m =
\mathcal{V}^{-1}\sum_{i}\sigma_{i}$. Taking the average over
different disorder configurations we may define the order
parameter of the system $M = [|m|]$, where $[\cdots]$ denotes
disorder averaging. Another physical parameter of interest is the
bond energy per spin that corresponds to the first term of the
Hamiltonian~(\ref{eq:1}), i.e. $e_{\rm J}=-
\mathcal{V}^{-1}\sum_{\langle i,j \rangle}\sigma_{i}\sigma_{j}$, and its
disorder average, defined hereafter as $E_{\rm J}=[e_{\rm J}]$.
Our analysis in the sequel will be mainly based on these three
thermodynamic quantities, as well as a relevant algorithmic
quantity, namely the number of primitive operations of the PR
algorithm, that is the number of relabels per spin $R$.

\begin{figure}[!t]
\centerline{
\includegraphics*[width=0.5\textwidth]{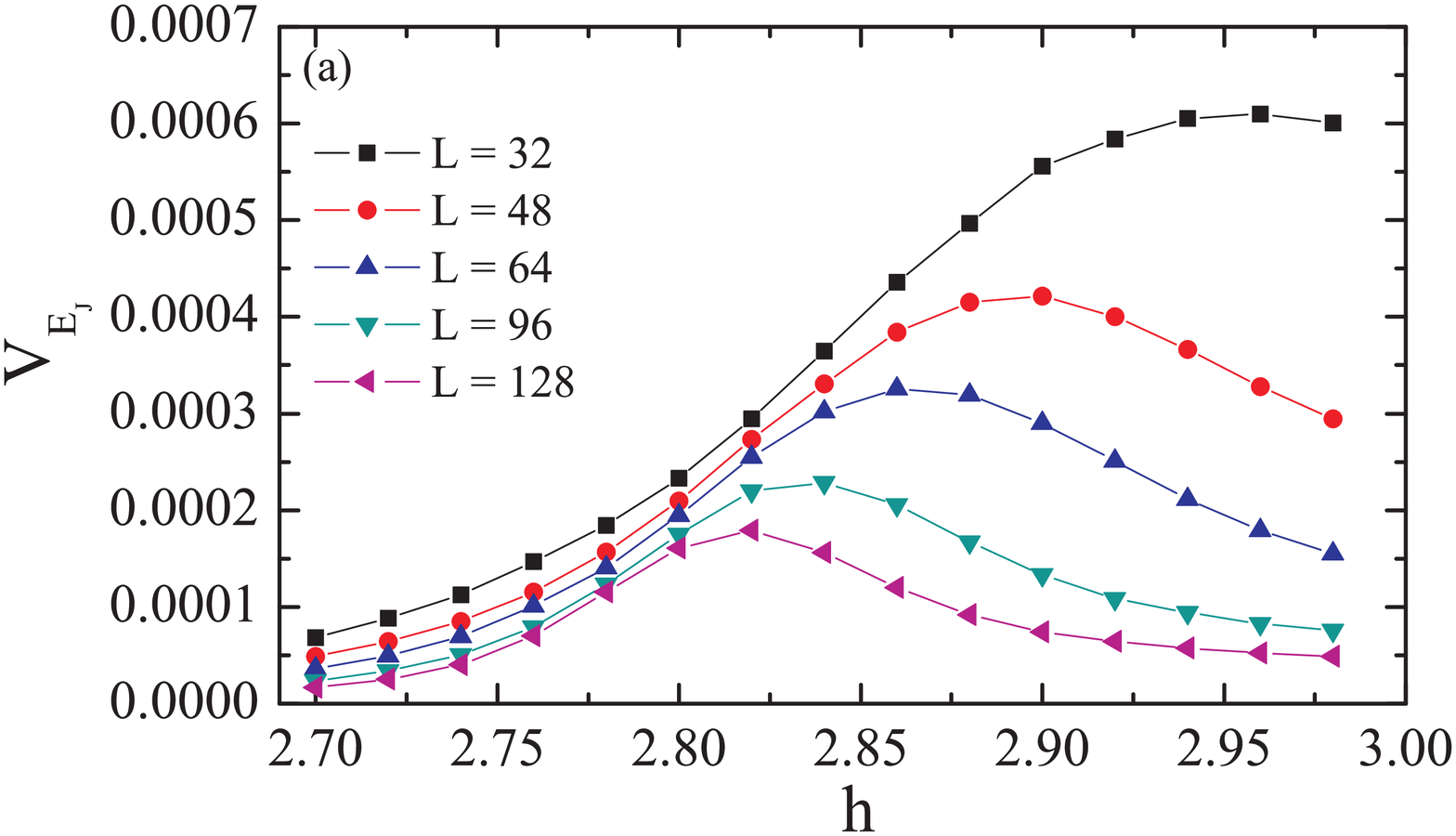}
\includegraphics*[width=0.49\textwidth]{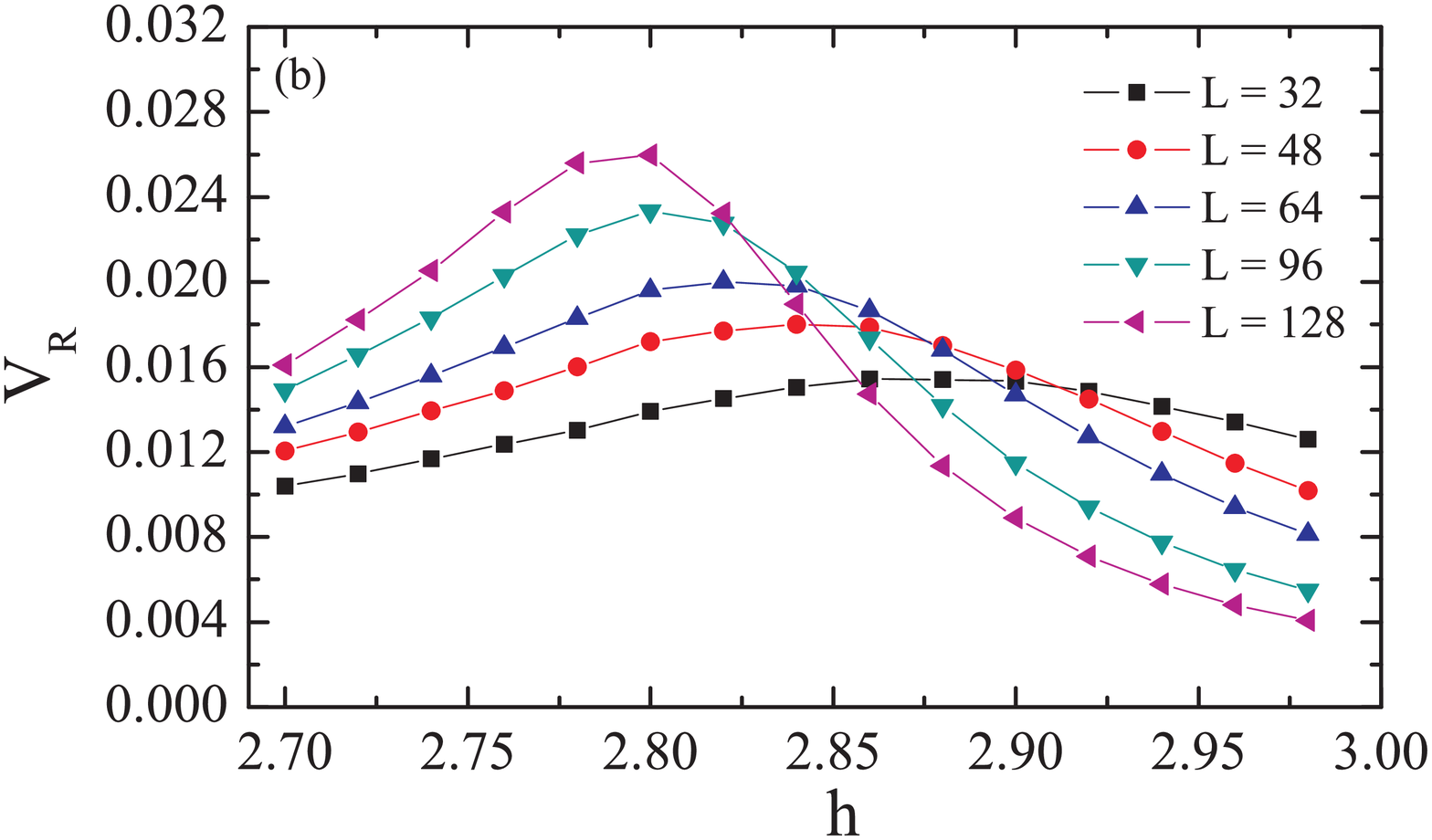}
}
\vspace{2mm}
\centerline{
\includegraphics*[width=0.5\textwidth]{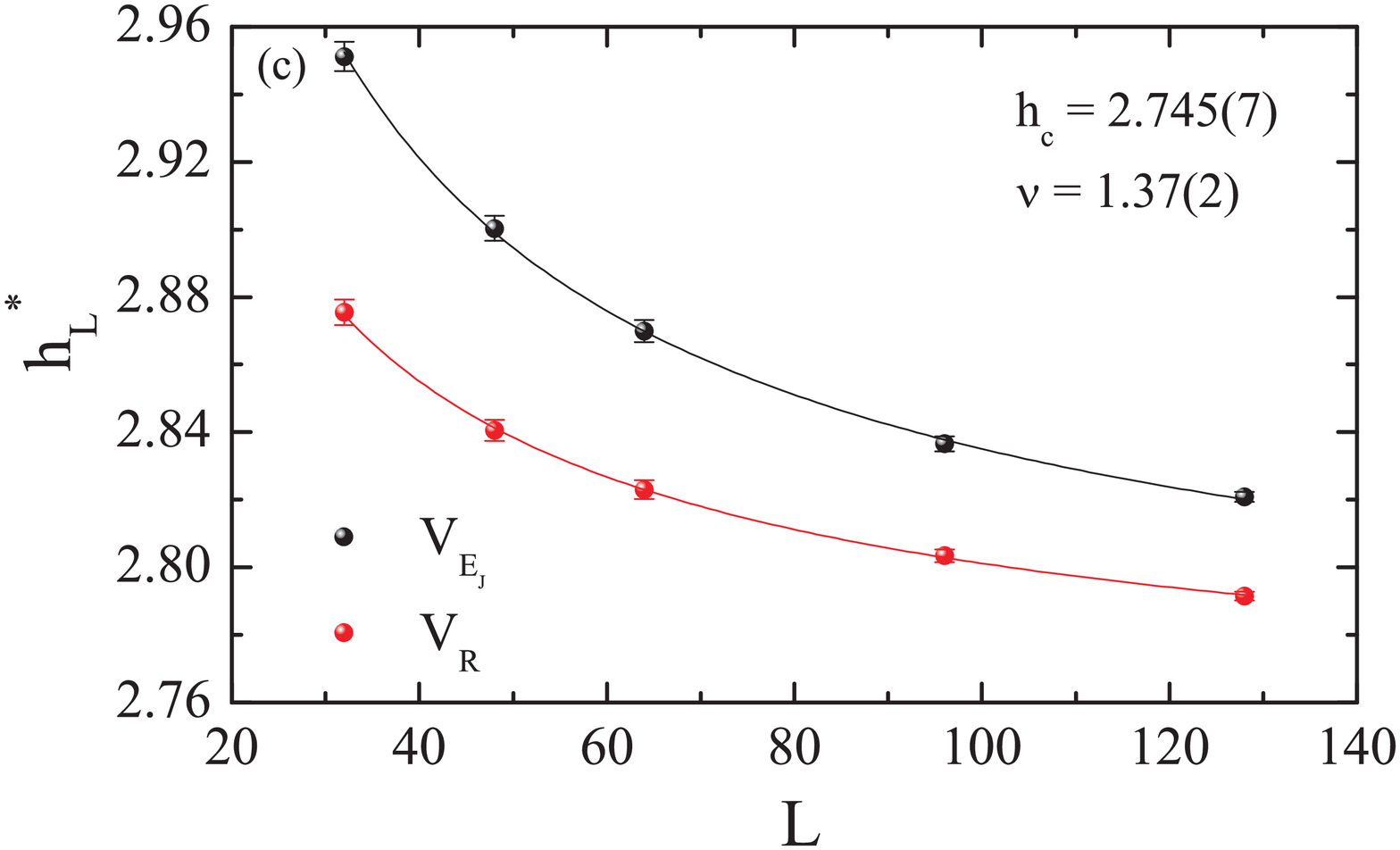}
}
\caption{\label{fig:2} (Color online) (a) Sample-to-sample
fluctuations of the bond energy $V_{E_{\rm J}}$ of the trimodal
RFIM as a function of the disorder strength for various lattice
sizes in the range $L=32-128$. Lines are simple guides to the eye.
(b) Same as in panel (a) but now the sample-to-sample fluctuations
of the number of primitive operations of the PR algorithm, $V_{\rm
R}$, are shown. (c) Simultaneous fitting of the form~(\ref{eq:7})
of the pseudo-critical disorder strengths $h_{L}^{\ast}$,
obtained from the peak positions of the fluctuations shown in
panels (a) and (b). The shared parameters of the three data sets
of the fit are the critical strength $h_{\rm c}$ and the
correlation length's exponent $\nu$.}
\end{figure}

At this point, let us start the presentation of our FSS approach
with figures~\ref{fig:2}~(a) and (b), where we plot the
sample-to-sample fluctuations over disorder of two quantities, of
physical and technical origin, for the case of the trimodal RFIM.
In particular, we plot the fluctuations of the bond energy $E_{\rm
J}$ [figure~\ref{fig:2}~(a)] and the number of primitive operations
of the PR algorithm [figure~\ref{fig:2}~(b)]. All these fluctuations
are plotted as a function of the disorder strength $h$ for the
complete lattice size-range $L=32-128$. It is clear that for every
lattice size $L$, these fluctuations appear to have a maximum
value at a certain value of $h$, denoted hereafter as $h_{L}^{\ast}$, that may be considered in the following as a suitable
pseudo-critical disorder strength. By fitting the data points
around the maximum first to a Gaussian, and subsequently to a
fourth-order polynomial, we have extracted the values of the
peak-locations ($h_{L}^{\ast}$) by taking the mean value via
the two fitting functions, as well as the corresponding error
bars. Using now these values for $h_{L}^{\ast}$ we consider in
the panel (c) of figure~\ref{fig:2} a simultaneous power-law fitting
attempt of the form
\begin{equation}
\label{eq:7} h_{L}^{\ast}=h_{\rm c}+bL^{-1/\nu},
\end{equation}
simultaneous meaning that the values of $h_{\rm c}$ and $\nu$ for
all data sets in the fitting procedure are shared during the fit.
The quality of the fit is fair enough, with a value of
$\chi^{2}$/dof of the order of $0.6$, where dof refers to the
degrees of freedom, and produces the estimates $h_{\rm
c}=2.745(7)$ and $\nu=1.37(2)$ for the critical disorder strength
and the correlation length's exponent, in agreement with recent
estimates in the literature~\cite{fytas}.

\begin{figure}[!t]
\centerline{
\includegraphics*[width=0.47\textwidth]{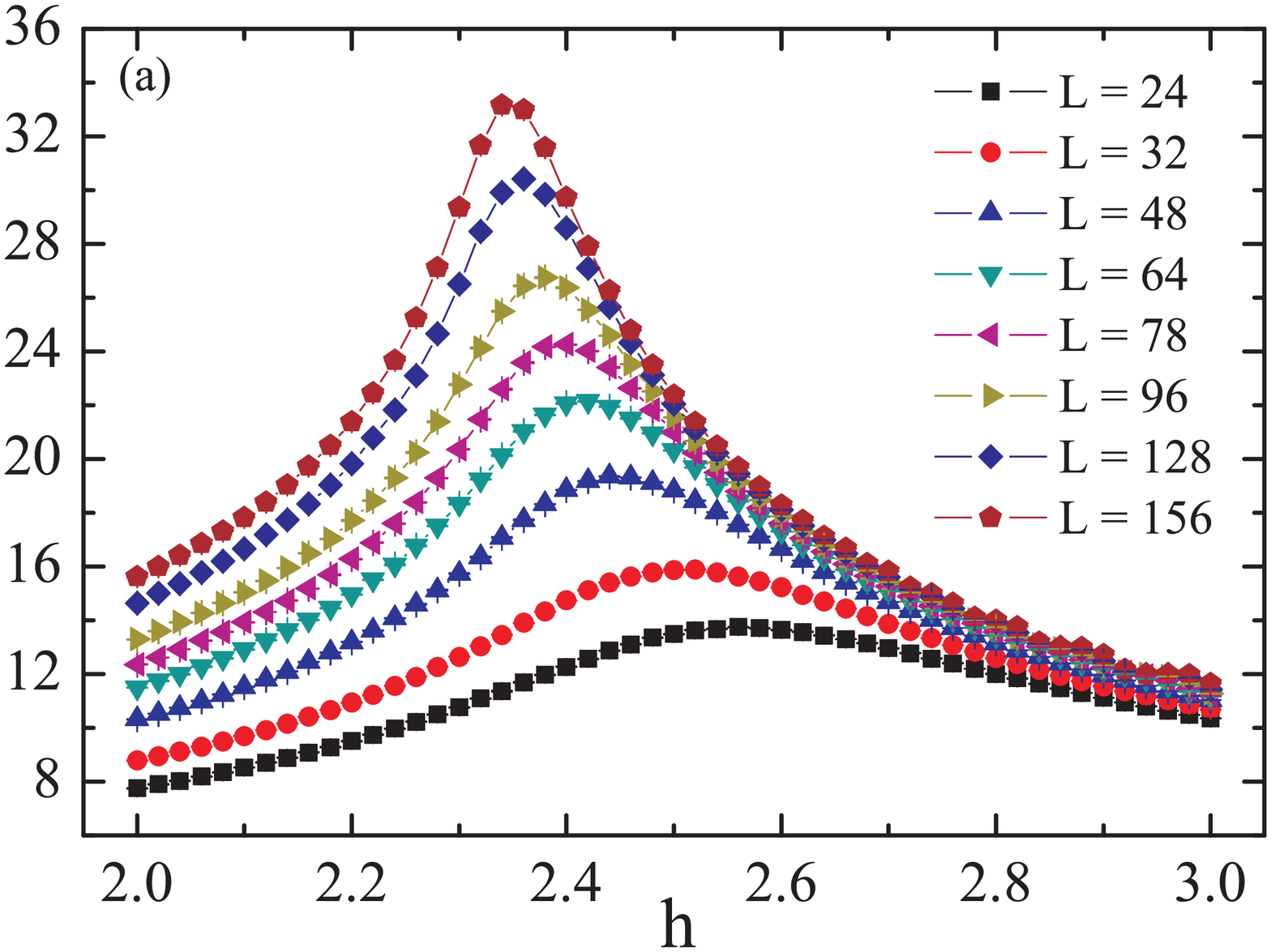}\hfill
\includegraphics*[width=0.47\textwidth]{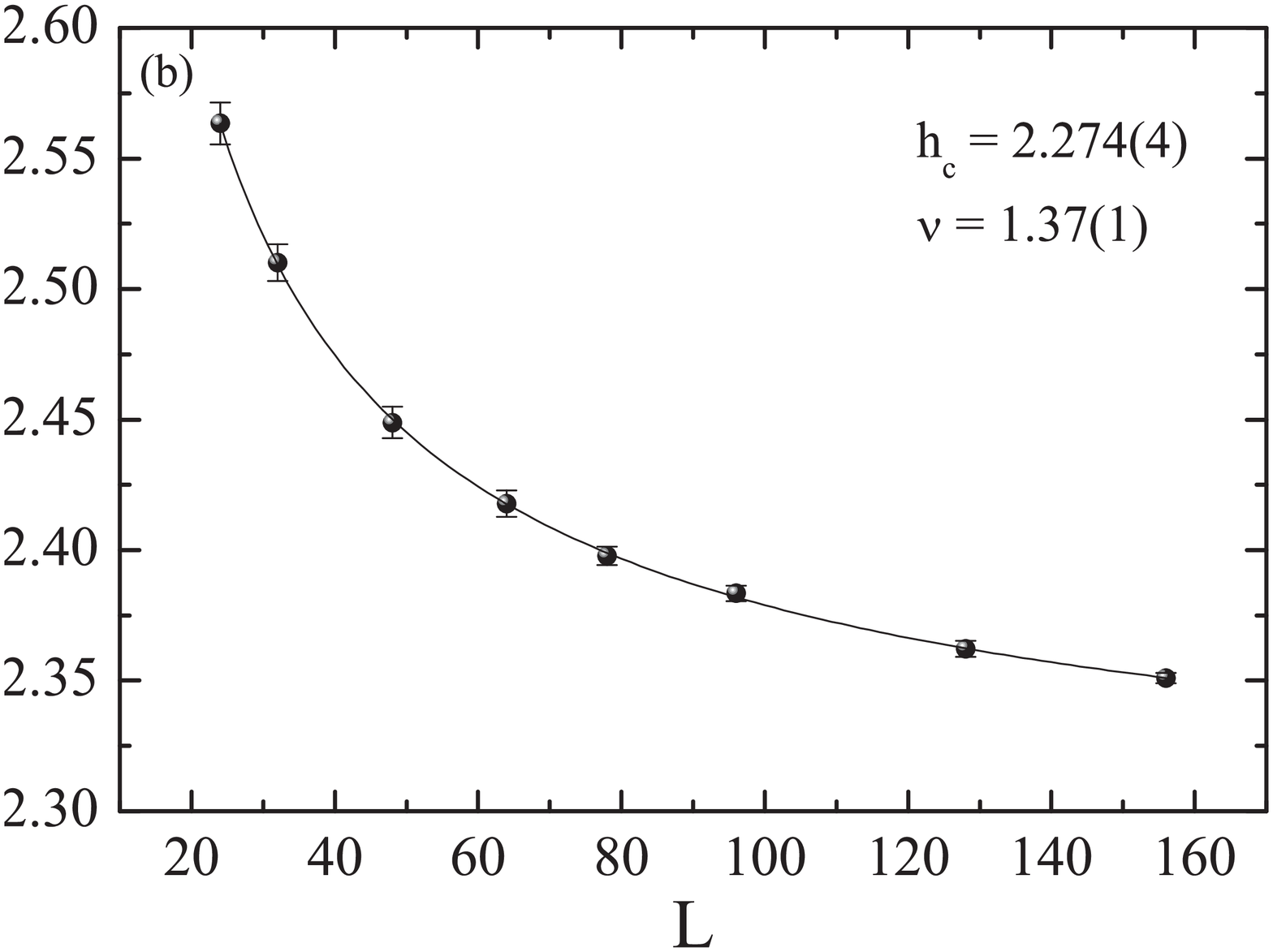}
}
\caption{\label{fig:3} (Color online) (a) The number of relabels
per spin $R$ of the Gaussian RFIM as a function of the disorder
strength for various lattice sizes in the range $L=24-156$. Lines
are simple guides to the eye. (b) Fitting of the form~(\ref{eq:7})
of the pseudo-critical disorder strengths $h_{L}^{\ast}$,
obtained from the peak positions of panel (a).}
\end{figure}

We now turn our discussion on the Gaussian RFIM. For this case we
show in figure~\ref{fig:3}~(a) the number of relabels per spin $R$ as
a function of the disorder strength for various lattice sizes in
the range $L=24-156$. Again, we observe that for every lattice
size $L$, $R$ has a maximum at a certain value of $h$, denoted as
before with $h_{L}^{\ast}$, that may be considered now as a
relevant pseudo-critical disorder strength. Following a similar
procedure, we extracted the values of the peak-locations ($h_{L}^{\ast}$) as well as the corresponding error bars, whose
shift-behavior is now plotted in panel (b) of figure~\ref{fig:3}.
The straight line is power-law fitting attempt of the same
form~(\ref{eq:7}) and the outcome for $h_{c}$ and $\nu$ is
$2.274(4)$ and $1.37(1)$, respectively. The quality of the fit is
also in this case good, with a value of $\chi^{2}$/dof of the
order of $0.4$.

A few comments on the scaling analysis are now in order: Having
simulated more than five lattice size-points in each case, we also
tried to perform the above analysis including higher-order scaling
corrections of the form $(1+b'L^{-\omega})$, where $\omega$ is the
well-known correction-to-scaling exponent, obtained very recently
to be $\omega=0.52(11)$ for this model~\cite{fytasPRL}, using the
scaling behavior of universal quantities. However, no improvement
has been observed in the quality of the fit of our data. On the
contrary, the corrected scaling assumption resulted in an unstable
fitting procedure with significantly large errors in the values of
the exponent $\nu$, the coefficient $b'$, as well as the exponent
$\omega$.

Our suggestion of choosing these newly defined pseudo-critical
disorder strengths $h_{L}^{\ast}$ as a proper measure for
performing FSS, closely follows the analogous considerations of
Hartmann and Young \cite{hartmann01} and Dukovski and
Machta \cite{machta03}, also for the Gaussian RFIM. The first
authors \cite{hartmann01} considered pseudo-critical disorder
strengths at the values of $h$ at which a specific-heat-like
quantity obtained by numerically differentiating the bond energy
with respect to $h$ attains its maximum. On the other hand, the
authors of reference~\cite{machta03} identified the pseudo-critical
points as those in the $H - h$ plane (with $H$ a uniform external
field), where three degenerate ground states of the system show
the largest discontinuities in the magnetization.

Respectively, Middleton and Fisher~\cite{middleton1} using similar
reasoning on the Gaussian RFIM, characterized the distribution of
the order parameter by the average over samples of the square of
the magnetization per spin and the root-mean-square
sample-to-sample variations of the square of the magnetization.
They identified a similar behavior to that of figures~\ref{fig:2}~(a)
and (b), i.e., with increasing $L$, the peak magnitude of this
quantity moved its location to smaller values of $h$, defining
another relevant pseudo-critical disorder strength. However, in
reference~\cite{middleton1} the authors were only interested in the
scaling behavior of the height of these peaks. The practice
followed in the current paper, employing the FSS behavior of the
peaks of the sample-to-sample fluctuations of several quantities
of physical ($M$ and $E_{\rm J}$) and technical ($R$) origin, was
inspired by the intriguing analysis of Efrat and
Schwartz~\cite{efrat06}. These authors, studying also the $d=3$
RFIM, showed that the behavior of the sample-to-sample
fluctuations in a disordered system may be turned into a useful
tool that can provide an independent measure to distinguish
between the ordered and disordered phases of the system. The
analysis of figures~\ref{fig:2}~(a) and (b) above verifies their
prediction, and the accuracy in the estimation of relevant phase
diagram features, like the critical field $h_{\rm c}$ and the
critical exponent $\nu$, turns out to be a clear test in favor of the
overall scheme.

Let us make at this point a small comment concerning the errors
inherent in these types of approximations. The errors induced in
the scheme based on the sample-to-sample fluctuations of
figures~\ref{fig:2}~(a), \ref{fig:2}~(b), or the primitive operations
of the PR algorithm shown in figure~\ref{fig:3}~(a), have their
origin in the application of some polynomial, or peak-like,
function in order to extract the relevant position of the maximum
in the $h$-axis. On the contrary, in similar definitions of
pseudo-critical points, such as through the use of some properly
defined specific-heat-like quantity at $T=0$~\cite{hartmann01},
one should numerically differentiate the data of the bond energy
$E_{\rm J}$, and then consider a smoothing function to locate the
position of the maximum. This scheme is subjected to two
successive fitting approximations, thus increasing the errors in
the estimation of the pseudo-critical points.

To summarize, in this section, we have investigated the matching
between the trimodal, $p=1/3$, RFIM and the Gaussian RFIM. Clearly
enough, our estimates for the critical exponent $\nu$ of both
models indicate an equivalence among both distributions within
error bars, justifying the original prediction of
Mattis~\cite{mattis}. Furthermore, we have suggested the values for
the critical field $h_{\rm c}$ which compare very well to the most
accurate estimations of the literature. For instance, the best
known estimate for the Gaussian RFIM is $h_{\rm c} =
2.27205$~\cite{fytasPRL}, very close to the value $2.274(4)$ of
the present work. This is also true for the reported values of the
correlation-length's exponent, as for the Gaussian RFIM, previous
high-accuracy estimates suggest a value of
$\nu=1.37$~\cite{fytasPRL,middleton1,hartmann01}. An interesting
aspect of our analysis that led to the above results was the
illustration that quantities related to the sample-to-sample
fluctuations of several quantities of the system or simply the,
originally technical, number of primitive operations of the PR
algorithm, constitute a useful alternative to investigate
criticality.

\section{Gaussian RFIM}
\label{sec:4}

In this last part of our work, we concentrate on the Gaussian
distribution, which is the most studied case in the literature of
the RFIM, and present further results on important aspects of its
critical behavior. As already mentioned above, we have performed
additional runs at the best-known value of the critical field,
that is the value $h_{c}=2.27205$~\cite{fytasPRL}. Thus, the data
and analysis of this section are based on extensive simulations
performed at this value of the critical field.

\begin{figure}[!b]
\centerline{
\includegraphics*[width=0.5\textwidth]{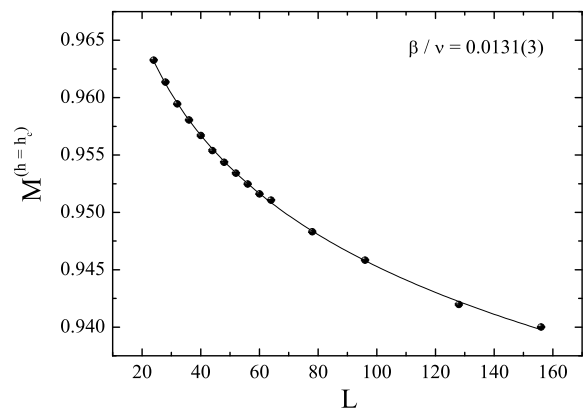}
}
\caption{\label{fig:4} FSS of the order parameter at the critical
field $h_{\rm c}$.}
\end{figure}

In principal, we are interested in the extraction of an accurate
estimate for the magnetic exponent ratio $\beta/\nu$, whose small
value casts some doubts on the order of the transition of the
RFIM. The route we follow here is via the scaling of the order
parameter $M$ at the critical field. This is shown in
figure~\ref{fig:4}, and the solid line is a power-law fitting of the
form $M^{(h=h_{\rm c})}\sim L^{-\beta/\nu}$. The resulting
estimate of the magnetic exponent ratio, given also in the figure,
is $\beta/\nu=0.0131(3)$, a rather small, but non-zero value, also in
agreement with some of the most accurate estimations in the
literature~\cite{middleton1}.

The next part of our FSS analysis concerns the controversial issue
of the specific heat of the RFIM. The specific heat of the RFIM
can be experimentally measured~\cite{jaccarino,jaccarino.2} and is, for sure,
of great theoretical importance. Yet, it is well known that it is
one of the most intricate thermodynamic quantities to deal with in
numerical simulations, even when it comes to pure systems. For the
RFIM, Monte Carlo methods at $T>0$ have been used to estimate the
value of its critical exponent $\alpha$, but were restricted to
rather small systems sizes and have also revealed many serious
problems, i.e., severe violations of selfaveraging~\cite{PS02,fytas06}. A better picture emerged throughout
the years from $T=0$ computations, proposing estimates of
$\alpha\approx 0$. However, even by using the same numerical
techniques, but different scaling approaches, some inconsistencies
were recorded in the literature. The most prominent was that
of reference \cite{hartmann01}, where a strongly negative value of the
critical exponent $\alpha$ was estimated. On the other hand,
experiments on random field and diluted antiferromagnetic systems
suggest a clear logarithmic divergence of the specific
heat~\cite{jaccarino,jaccarino.2}.

\begin{figure}[!t]
\centerline{
\includegraphics*[width=0.5\textwidth]{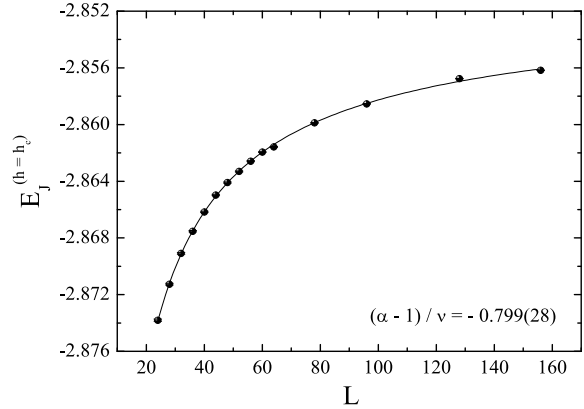}
}
\caption{\label{fig:5} FSS behavior of the bond part of the energy
density at the critical field $h_{\rm c}$. The line is a fitting
of the form~(\ref{eq:8}).}
\end{figure}

In general, one expects that the finite-temperature definition of
the specific heat $C$ can be extended to $T=0$, with the second
derivative of $\langle E\rangle$ with respect to temperature being
replaced by the second derivative of the ground-state energy
density $E_{\rm gs}$ with respect to the random field
$h$~\cite{middleton1,hartmann01}. The first derivative $\partial
E_{\rm gs}/\partial J$ is the bond energy $E_{\rm J}$, already
defined above. The general FSS form assumed is that the singular
part of the specific heat $C_{\rm s}$ behaves as $C_{\rm s}\sim
L^{\alpha/\nu}\tilde{C}\left [(h-h_{\rm c})L^{1/\nu}\right]$.
Thus, one may estimate $\alpha$ by studying the behavior of
$E_{\rm J}$ at $h = h_{\rm c}$~\cite{middleton1}. The computation
from the behavior of $E_{\rm J}$ is based on integrating the above
scaling equation up to $h_{\rm c}$, which gives a dependence
\begin{equation}
\label{eq:8} E_{\rm J}^{(h=h_{\rm
c})}=c_{1}+c_{2}L^{(\alpha-1)/\nu},
\end{equation}
with $c_{i}$ constants.
Alternatively, following the prescription of~\cite{hartmann01},
one may calculate the second derivative by finite differences of
$E_{\rm J}(h)$ for values of $h$ near $h_{\rm c}$ and determine
$\alpha$ by fitting to the maximum of the peaks in $C_{\rm s}$,
which occur at $h_{L}^{\ast}-h_{\rm c}\approx L^{-1/\nu}$.
However, as already noted in~\cite{middleton1}, this latter
approach may be more strongly affected by finite-size corrections,
since the peaks in $C_{\rm s}$ found by numerical differentiation
are somewhat above $h_{\rm c}$, and furthermore it is
computationally more demanding, since one must have the values of
$E_{\rm J}$ in a wide and very dense range of $h$-values.

In the present case, where the critical value $h_{\rm c}$ is known
with good accuracy, the first approach seems to be more suitable
to follow. The numerical data of the critical bond energy and the
relevant scaling analysis are presented in figure~\ref{fig:5}. The
solid line is a power-law fitting of the form~(\ref{eq:8}) and the
estimate for the exponent ratio $(\alpha-1)/\nu$ is $-0.799(28)$,
as also given in the figure. Using now our estimate $\nu=1.37(1)$,
we calculate the critical exponent $\alpha$ of the specific heat,
resulting in an estimate $\alpha=-0.095(37)$, which is fairly
compatible to the experimental scenario of a logarithmic
divergence ($\alpha=0$)~\cite{jaccarino,jaccarino.2}.

Following the discussion above in section~\ref{sec:1}, our numerical
studies of disordered systems are carried out near their critical
points using finite samples; each sample is a particular random
realization of the quenched disorder. A measurement of a
thermodynamic property, say $Z$, yields a different value for
every sample. In an ensemble of disordered samples of linear size
$L$, the values of $Z$ are distributed according to a probability
distribution. The behavior of this distribution is directly
related to the issue of self-averaging. In particular, by studying
the behavior of the width of this distribution, one may
qualitatively address the issue of self-averaging, as has already been
stressed by previous authors~\cite{WD95,AH96,WD98}. In general, we
characterize the distribution by its average $[Z]$ and also by the
relative variance
\begin{equation}
\label{eq:9}
R_{Z}=\frac{V_{Z}}{[Z]^{2}}=\frac{[Z^{2}]-[Z]^{2}}{[Z]^{2}}\,,
\end{equation}
that we employ here to investigate the self-averaging properties
of the RFIM.

\begin{figure}[!t]
\centerline{
\includegraphics*[width=0.52\textwidth]{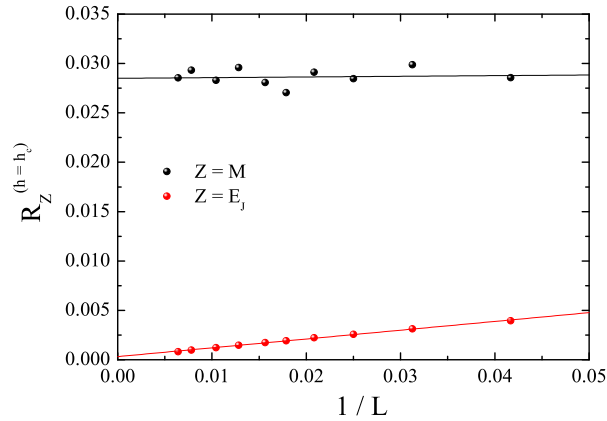}
}
\caption{\label{fig:6} (Color online) Illustration of the
self-averaging properties of the model in terms of the
magnetization and bond energy. The lines are linear extrapolations
to the infinite-limit size.}
\end{figure}

In particular, we study the behavior of the ratio $R_{Z}$, in the
framework of the two main quantities typically used, the order
parameter $M$ and the bond energy $E_{\rm J}$ of the model. In
figure~\ref{fig:6} we plot the ratio $R_{Z}$, estimated at the
critical field $h_{\rm c}$, for both quantities, as a function of
the inverse linear size. The solid lines are simple linear
extrapolations to the infinite-limit size $L\rightarrow \infty$.
As it is straightforward from the extrapolations, the
order-parameter that carries the effect of the disorder~--- we
remind here that the random field couples to the local spins~--- is
a strongly non-self-averaging quantity. On the other hand, as
expected, the bond energy restores self-averaging in the
thermodynamic limit.

\begin{figure}[!b]
\centerline{
\includegraphics*[width=0.49\textwidth]{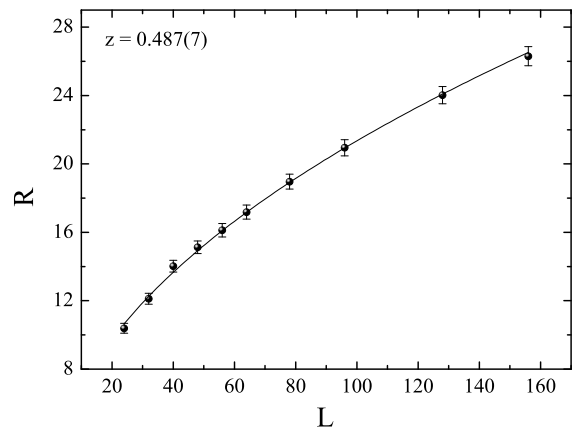}
}
\caption{\label{fig:7} Estimation of the critical slowing down
exponent $z$ of the PR algorithm via the FSS behavior of the
number of relabels per spin at the critical field $h_{\rm c}$.}
\end{figure}

Closing, we present some computational aspects of the implemented
PR algorithm and its performance on the study of the Gaussian
RFIM. Although its generic implementation has a polynomial time
bound, its actual performance depends on the order in which
operations are performed and which heuristics are used to maintain
auxiliary fields for the algorithm. Even within this polynomial
time bound, there is a power-law critical slowing down of the PR
algorithm at the zero-temperature
transition~\cite{ogielski85,middleton1}. This critical slowing
down is certainly reminiscent of the slowing down seen in local
algorithms for statistical mechanics at finite temperature, such
as Metropolis, and even for cluster algorithms~\cite{landau}. In
fact, Ogielski was the first to note that the PR algorithm takes
more time to find the ground state near the transition in three
dimensions from the ferromagnetic to paramagnetic
phase~\cite{ogielski85}.
This has already been qualitatively seen in figure~\ref{fig:3}~(a),
where, indeed, the number of primitive operations $R$ of the PR
algorithm is maximized in the suitably defined pseudo-critical
fields $h_{L}^{\ast}$. Assuming the standard scaling of the
form $R \approx L^{z} w\left[(h-h_{\rm c})^{-1/\nu}L\right]$,
where the dynamic exponent $z$ describes the divergence in the
running time at $h=h_{\rm c}$, and $w(x)\sim x^{-z}$ at large $x$
and $w(x)\sim |x|^{-z}\ln{|x|}$ as $x\rightarrow -\infty$, to be
consistent with convergence to constant $R$ for $h>h_{\rm c}$ and
$R\sim \ln{L}$ for small $h$. Our fitting attempt of this scaling
form is plotted in figure~\ref{fig:7} and the obtained estimate for
the dynamic critical exponent $z$ is $0.487(7)$.

\clearpage

\begin{figure}[!t]
\centerline{
\includegraphics*[width=0.51\textwidth]{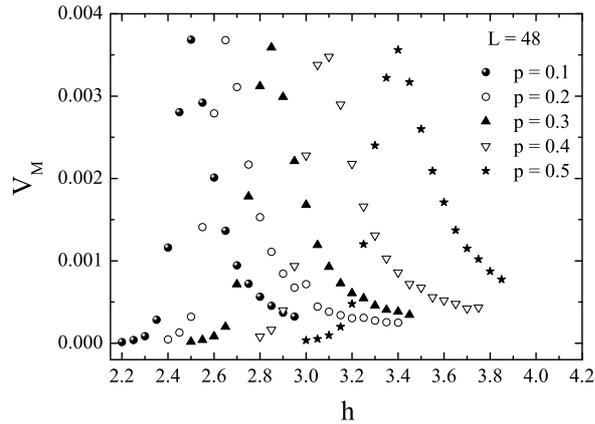}
}
\caption{\label{fig:8} Sample-to-sample fluctuations of the order
parameter of an $L=48$ trimodal RFIM with varying probability $p$
as a function of the external random field $h$. A clear shift
behavior is observed.}
\end{figure}

\section{Summary and outlook}
\label{sec:5}

To summarize, we have investigated the ground-state criticality of
the $d=3$ RFIM with two types of the random-field distribution,
namely a uniform trimodal and a Gaussian distribution. In
particular, we have estimated for both cases the critical disorder
strength $h_{\rm c}$ and the critical exponent $\nu$ of the
correlation length. These values, compare well enough to the most
accurate estimates of the literature, with the values of $\nu$
placing the trimodal ($p=1/3$) RFIM into the universality class of
the Gaussian model, thus verifying a scenario suggested many years
ago by Mattis~\cite{mattis}. Technically, our effort became
feasible through the implementation of a modified version of the
PR algorithm that enabled us to simulate very large system sizes,
up to $156^{3}$ spins, and disorder ensembles of the order of up
to $200\times 10^{3}$, for several values of the random-field
strength.

On physical grounds, we have implemented a FSS approach based on
the sample-to-sample fluctuations of various quantities of
physical and technical origin and the primitive operations of the
PR algorithm. The outcome of this analysis indicated that the
fluctuations of the system may be used as an alternative
successful approach to criticality, paving the way to even more
sophisticated studies of disordered systems under this
perspective. Furthermore, we have provided high-accuracy estimates
for the controversial issues of the magnetic-exponent ratio of the
order parameter $\beta/\nu$ and the critical exponent $\alpha$ of
the specific heat. In particular, the magnetic exponent ratio
$\beta/\nu$ was found to be very small, but clearly non zero,
ruling out the possibility of a first-order phase transition,
whereas the exponent $\alpha$ was found to be compatible with
zero, in agreement with the experiments~\cite{jaccarino,jaccarino.2}.
Particular interest has been paid to the self-averaging properties
of the model, by studying the infinite-limit size extrapolation of
energy- and order-parameter-based noise to signal ratios, as well
as the critical slowing down aspects of the PR algorithm.

A future challenge emerging out of the current work, is the
production of the phase diagram of the trimodal RFIM in the
$h_{\rm c} - p$ plane and the verification, or challenge, of the
originally proposed for $p=1/3$ universal behavior of the trimodal
and Gaussian models in higher dimensions, below the upper critical
dimension of the RFIM $d_{\rm u}$. Preliminary simulations for
various values of the probability $p$ in the spectrum $0.1-0.5$ of
the trimodal RFIM indicate a smooth scaling behavior of the
sample-to-sample fluctuations of the order parameter, as
illustrated in figure~\ref{fig:8} for a system size of $L=48$ and
ensembles of the order of $\mathcal{N}_{\rm s}=5\times 10^3$
realizations. We expect this task and analysis to bring forward
new results on the RFIM that will be useful to the community of
disordered systems.

\clearpage

\ukrainianpart
\title{Модель Ізінга у випадковому полі: моделювання при нульовій температурі}

\author{П.Е. Теодоракіс\refaddr{ad1}, Н.Г. Фітас\refaddr{ad2}}
\addresses{
\addr{ad1} Відділ хімічної інженерії, Емпіріал Коледж Лондон,
SW7 2AZ, Лондон, Великобританія
\addr{ad2} Центр прикладних математичних досліджень, Університет м. Ковентрі,
Ковентрі, CV1 5FB, Великобританія}

\makeukrtitle

\begin{abstract}
\tolerance=3000%
Застосовуючи комп'ютерні симуляції при нульовій температурі, ми висвітлюємо деякі аспекти критичної поведінки тривимірної ($d=3$)
моделі Ізінга у випадковому полі. Ми розглядаємо дві версії моделі, що відрізняються  розподілом випадкового поля,
а саме, гаусову та тримодову моделі Ізінга у випадковому полі з однаковими вагами.
Застосовуючи обчислювальний підхід, що ставить у відповідність основному стану системи
проблему оптимізації максимуму потоку на мережі, ми використовуємо найсучаснішу версію
алгоритму проштовхування потоку і моделюємо великі ансамблі  випадкових реалізацій
моделей для широкої області значень випадкового поля і розмірів системи $\mathcal{V}=L\times L\times L$,
де $L$ позначає лінійний розмір гратки і $L_{\rm max}=156$. Використовуючи в якості скінчено-вимірних
мір флуктуації різних величин фізичного і технічного походження, виміряних для різних зразків,
і примітивні операції алгоритму проштовхування потоку, ми пропонуємо для обох типів розподілу
оцінки критичного поля $h_{\rm c}$ і критичного показника кореляційної довжини $\nu$. Отримане
значення цього показника чітко вказує на те, що обидві моделі
належать до одного класу універсальності. Додаткові симуляції гаусової моделі Ізінга у випадковому полі при добре
відомому значенні критичного поля забезпечують відношення магнітних індексів
$\beta/\nu$ з високою точністю і прояснюють контроверсійну проблему критичного індекса $\alpha$ питомої теплоємності.
Накінець, ми обговорюємо нескінченнорозмірну
екстраполяцію енергії і базованого на параметрі порядку шуму до сигнальних коефіцієнтів,
пов'язаних з властивостями самоусереднення моделі, а також аспекти критичного сповільнення алгоритму.
\keywords модель Ізінга у випадковому полі, скінченнорозмірний скейлінг, теорія графів
\end{abstract}

\end{document}